\documentclass[iop]{emulateapj}
\usepackage{graphicx}

\usepackage{color}

\usepackage{epstopdf}
\usepackage{amsmath}

\def\ltsima{$\; \buildrel < \over \sim \;$}
\def\simlt{\lower.5ex\hbox{\ltsima}}
\def\gtsima{$\; \buildrel > \over \sim \;$}
\def\simgt{\lower.5ex\hbox{\gtsima}}

\newcommand\lsim{\mathrel{\rlap{\lower4pt\hbox{\hskip1pt$\sim$}}
\raise1pt\hbox{$<$}}}
\newcommand\gsim{\mathrel{\rlap{\lower4pt\hbox{\hskip1pt$\sim$}}
\raise1pt\hbox{$>$}}}
  
\usepackage{color}

\shorttitle{It's a planetary feast}
\shortauthors{Qureshi et al.}

\begin{document}

\title{Signature of Planetary Mergers on Stellar Spins }

\author{Ahmed Qureshi\altaffilmark{1} Smadar Naoz\altaffilmark{1,2},  Evgenya L.~ Shkolnik\altaffilmark{3}\\}
\affil{$^1$Department of Physics and Astronomy, University of California, Los Angeles, CA 90095, USA}
\affil{$^2$Mani L. Bhaumik Institute for Theoretical Physics, Department of Physics and Astronomy, UCLA, Los Angeles,
CA 90095}
\affil{$^3$School of Earth and Space Exploration, Arizona State University, Tempe, AZ 85281, USA}

\email{ahmed.qureshi@ucla.edu \\ snaoz@astro.ucla.edu }

\begin{abstract}
One of the predictions of high eccentricity planetary migration is that many planets will end up plunging into their host stars. We investigate the consequence of planetary mergers on their stellar hosts' spin-period. Energy and angular momentum conservation yield that a planet consumption by a star will spin-up the star. We find that our proof-of-concept calculations align with the observed bifurcation in the stellar spin-period in young clusters. For example, after a Sun-like star has eaten a Jupiter-mass planet it will spin up by $\sim 60\%$ (i.e., spin-period is reduced by $\sim 60\%$), causing an apparent gap in the stellar spin period, between stars that consumed a planet and those that did not. The spun-up star will later spin down due to magnetic braking, consistent with the disappearance of this bifurcation in clusters ($\gsim 300$Myr) . The agreement between the calculations presented here and the observed spin-period color diagram of stars in young clusters, provides circumstantial evidence that planetary accretion onto their host stars is a generic feature of planetary-system evolution.
\end{abstract}
\keywords{exoplanets, stars, young clusters}

\maketitle

\section{Introduction}

Recent observations  showed that many short-period exoplanets are found nearly at, or even interior to, their Roche limit  \citep[see figure 1 in][]{Jackson+17}, implying that these planets will be consumed by their host star. In the process of spiraling into a star, star-planet tidal interactions tend to spin up the star \citep[e.g.,][]{Dobbs-Dixon+04,Jackson+08,Jackson+09,Lanza+10,Brown+11,Bolmont+12}.
Further, it was proposed that a shortage of close-in planets around fast rotators \citep[e.g.,][]{McQuillan+13} can be attributed to the tidal merger of planets onto a star \citep[e.g.,][]{Teitler+14,Lanza+14}.

Driving a planet on a nearly radial orbit seems to be one of the natural consequences of the eccentric Kozai-Lidov mechanism \citep[see for recent review][]{Naoz16}. In this process, a far away companion can induce large planetary orbit's eccentricity, and plunge it into the star \citep[e.g.,][]{Guillochon+11,Naoz+12bin,Li+13,Rice15,Valsecchi+14Edge,Petrovich15He,Petrovich152p,Stephan+17}.  Smaller and moderate eccentricities are expected from  planet-planet interactions, but may still result in many planets plunging into the star \citep[e.g.,][]{Naoz11,Antonini+16,Petrovich+16,Hamers+17}. Here we show that as a planet falls onto a star, it deposits its energy, angular momentum and mass into the star, causing the star to spin up.

Stellar rotation is attributed to a combination of both  
stellar mass and evolutionary state. Sun-like stars  spin down by losing angular momentum to magnetized stellar winds, otherwise known as magnetic braking, during the main sequence stage  \citep[e.g.,][]{Parker58,Schatzman+62,Weber+67,Mestel+68}. 
Therefore, the stellar rotation-period or the rotational velocity $v_{\rm rot}\sin i$ 
is frequently used as a proxy for stellar ages  \citep[e.g.,][]{Barnes03L,Barnes03,Barnes+07, Mamajek+08, Meibom+09, Meibom+11,James+10,Mamajek+08, Meibom+09, Meibom+11,vanSaders+13,vanSaders+16}.

\citet{Mamajek+08} showed that {the spin-period of stars in the Pleiades open cluster  (age $\approx 130$~Myr) exhibits} a bifurcation for the same effective temperature (B-V) values. One group of stars are  fast rotators {\bf (with spin-periods of $1-2$ days)} and another one, slower rotators (with a spin-periods of $3-9$ days). \citet{Mamajek+08} showed that the latter group's spin-period traces the spin fit models adopted from \citet{Barnes07}.
This behavior is also observed in other young clusters such as M35 and M34 \citep[100~Myr and 240~Myr respectively, e.g.][]{Meibom+09,James+10}\footnote{See table \ref{table} for open cluster age estimations. }. Moreover, it seems that fast rotators have a dearth of close-in planets around them \citep[e.g.,][]{McQuillan+13,Lanza+14}. Observations suggest that this bifurcation is suppressed for older clusters such as the Hyades ($\approx 625$~Myr) and M~48 ($\approx 380$~Myr) \citep{Saar+99,Pizzolato+03,Mamajek+08,Meibom+09,Nardiello+15}.


	One interpretation for this division in rotation periods is that the fastest rotators have an outer convective magnetic field zone that shears the interior radiative zone and causes the gap in the rotation \citep[e.g.][]{Barnes03L,Barnes03,Meibom+09,James+10}. In other words, this interpretation suggests that  fast rotators, possess
only a convective field. Thus, they are inefficient in depleting their spin angular momentum. 
Later, \citet{vanSaders+16}, using evolutionary modeling, suggested that this gap is a result of a weaker magnetic braking process. Their models were able to reproduce both the asteroseismic and the cluster data. Recently, \citet{Somers+17}, analyzed the radii of single stars in the Pleiades and showed that inflated stars have a shorter spin period. Their statistical analysis included the inflation of young stars by magnetic activity and/or star-spots.  Furthermore, stellar evolution models of zero-main sequence radii contraction were able to produce consistent results with observations of rotation period in the star-forming regions and young open clusters
\citep[e.g.,][]{Gallet+13,Gallet+15}. {Recently, a new model for stellar} spin-down by \citet{Garraffo+18} was suggested, taking into account the   stellar surface magnetic field configuration as a plausible explanation for the observed  bimodal spin period distribution.


Observations of the young open clusters such as h Persei, \citep[$\sim13$~Myr, e.g.,][]{Moraux+13}, NGC 2264 \citep[$\sim 2$~Myr, e.g.,][]{Kearns+97} and NGC 2362 \citep[$\sim 5$~Myr, e.g.,][]{Irwin+08}, 
  give a glimpse to the birth rotational period distribution of stars. Unlike the clusters mentioned above, these extremely young clusters do not show a clear bifurcation signature, but instead a nearly uniform
distribution. For slightly older clusters ($\gsim 100$~Myr) a clear split between fast and slow rotators emerges, with the slow rotators spin periods fit the magnetic braking stellar evolution (see below, Section \ref{sec:MB} and Figure \ref{fig:Obs}).  For much older clusters ($\gsim 400$~Myr) the bifurcation disappears. The question then arise, what is the underlining mechanism that produces or maintains only the fast rotator population for a period of few hundreds of million years. The aforementioned magnetic models might be at play.   

Here we offer an alternative scenario: an increase in stellar rotation of a star due to the consumption of a Jupiter mass planet. A planet may plunge into the star, for example, due to high eccentricity migration. Thus, the star will absorb both the mass of the planet and the planet's orbital angular momentum, and will cause the star to spin up. 

In this model, the young clusters represent the birth population, {before giant planets formed} \citep[since,  typically expected to take place on a fews to $\sim10$~Myr, e.g.,][]{Pollack+96}. Subsequently, magnetic braking will drive to slow down the stars. 
As dynamical processes take place on the order of $\sim 10-100$~Myr, stars consume planets and thus spin up.

We note that the {detailed process} at which a planet accretes onto the star is complicated \citep[e.g.,][]{Metzger+12,Metzger+17,Pejcha+16,Dosopoulou+17,Ginzburg+17}. However, our calculations are independent of the process and depend only on the result, because {\bf we consider angular momentum conservation,} associated with the merger.  
We compare our calculations with several observed open cluster period-color diagrams and show that a planet consumed by a star at about $100$~Myr fits the observations.

	The paper is organized as follow. We begin with considering the effects on the stellar spins (Section \ref{sec:spin}). In particular we consider magnetic braking (Section \ref{sec:MB}), angular momentum conservation (Section \ref{sec:L}) and energy arguments (Section \ref{sec:energy}). We then continue with a description of the consequences of consumption of a planet on the stellar spin-period (Section \ref{sec:con}).We finally offer our discussion in Section \ref{sec:Diss}.

\section{Effects on the stellar spin}\label{sec:spin}

\subsection{Magnetic Braking}\label{sec:MB}

As mentioned, Sun-like stars undergo spin-down due to magnetic braking \citep[e.g.,][]{Parker58,Schatzman+62,Weber+67,Mestel+68}.
The spin loss rate {is evaluated as:} 
\begin{equation}\label{eq:MB}
   \dot{\mathbf{\Omega}} = -\alpha_{mb}\Omega^2\mathbf{\Omega} \ ,
\end{equation}
\citep[e.g.,][]{Dobbs-Dixon+04}, where 
\begin{equation} 
\alpha_{mb} = \left\{
\begin{array}{rl}
 1.5\times10^{-14}~{\rm yr} & {\rm G~stars} \\
 1.5\times10^{-15}~{\rm yr} & {\rm F~stars}
  \end{array} \right.
   \end{equation}
   {We use {\tt SSE} (single-star evolution) code \citep[e.g.,][]{Hurley+00} with the magnetic braking from Equation (\ref{eq:MB}) to calculate the spin evolution of the stars as a function of time for stellar masses between $0.6-1.8$~M$_\odot$. 
   Note that by implementing Equation (\ref{eq:MB}) in SSE we also include  the nominal dependency of the mass of the stellar envelope \citep[e.g.,][]{Hurley+00}, which result in a different time evolution. For example, the magnetic braking from \citet{{Hurley+00}} underestimates the spin period of a Sun-like star by a factor of $5$ at $\sim 5$~Gyr, while the \citet{Dobbs-Dixon+04} recipe yields a closer value for the Sun's spin rate. We take  initial spins from {\tt SSE}, and they range from $10.4$~day -- $0.8$~day for masses $0.6-1.8$~M$_\odot$, respectively.}


\subsection{Angular Momentum arguments}\label{sec:L}

\begin{figure*}
\centering
\includegraphics[width=\linewidth]{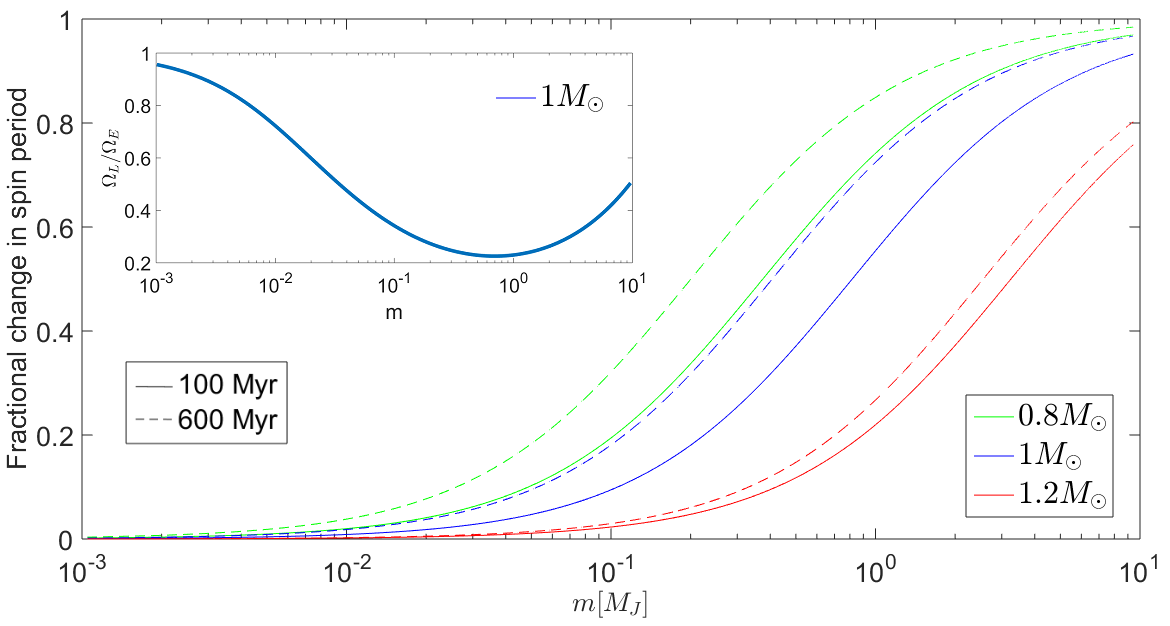}
\caption{ The spin-period absolute percentage change defined as: $|P_{\Omega,i}-P_{\Omega,f}|/P_{\Omega,i}$ as a function of the planet mass. The solid lines depict a merger that took place after $100$~Myrs, while the dashed lines represent a merger after $600$~Myrs of stellar evolution. The planet was assumed to plunge in from a $5$~au distance (as noted above the actual initial distance does not change significantly  the results). We consider three representative stellar masses $0.8,1$ and $1.2$~M$_\odot$, green, blue and red respectively. respectively. {In the inset, the ratio of the post-spin for angular momentum over conservation of energy is plotted versus the mass of the planet in terms $M_J$ for $1$~M$_\odot$ for a collision at $100$~Myrs. }    }\label{fig:timeplot}
\end{figure*}

{During the final step of high eccentricity migration, when the planet plunges in, we assume angular momentum conservation. We note that during the dynamical evolution the angular momentum of the inner orbit is not necessarily conserved as an inclined companion can exchange angular momentum with the inner planet \citep[e.g.,][]{Naoz11,Naoz+11sec}. However, {during the final plunge,  the inner orbit decouples from the outer orbit,} and thus it can be characterized with angular momentum conservation \citep[e.g.,][figure 3]{NF}. The equation in this case is: }
\begin{equation}\label{eq:Li}
  {\bf L}_i=I_s{\bf \Omega}_s+I_p{\bf \Omega}_p+{\bf L}_{\rm orb} \ ,
\end{equation}
where $\Omega_s$ ($\Omega_p$) is the star's (planet's) spin rate, where the star's and the planet's moments of inertia are $I_s=0.08M R^2$ and $I_p=0.26 m r^2 $, respectively \citep[e.g.,][]{Egg+01}, where $R$ ($r$)  radius of  star (planet). {We note that the radius of the star is assumed to stay constant post-consumption.}  Subscript ``i" denotes  denotes the initial (pre-consumption) state. 
 	The magnitude of the orbital angular momentum $L_{\rm orb}$ for an orbit with a semi-major axis $a$ and eccentricity $e$ is given by
\begin{eqnarray}\label{eq:Lorb}
 L_{\rm orb}&=&\frac{Mm}{M+m}\sqrt{G(M+m)a (1-e^2)} \\ &\approx&  \frac{Mm}{M+m}\sqrt{2 G(M+m) R_{\rm Roche} } \ ,\nonumber
\end{eqnarray}
where  $M$ and $m$ are the masses of the star and the planet respectively. {For plausible $\Omega_p$, the angular momentum associated with  planetary spin can be neglected compared to $L_{\rm orb}$ and $I_s \Omega_s$. Thus, Equation (\ref{eq:Li}) can be approximated as:}
\begin{equation}\label{eq:Li2}
  {\bf L}_i \approx I_s{\bf \Omega}_s+{\bf L}_{\rm orb} \ ,
\end{equation}

In the last transition in Equation (\ref{eq:Lorb}) we assumed high eccentricity so $e\to1$ and thus, $a(1-e^2) \approx 2 a(1-e)\approx 2 R_{\rm Roche}$, where $R_{\rm Roche}$ is the Roche limit given by  
\begin{equation}\label{eq:Roche}
  R_{\rm Roche} \sim \eta r \left(\frac{m}{M+m} \right)^{-1/3} \ ,
\end{equation}
where $\eta$ is a numerical parameter of the order of unity. We adopt $\eta=1.6$ and note that changing the value of $\eta$ does not qualitatively change the dynamical nature of the system but rather the efficacy of disruption of planets \citep[e.g.,][]{Petrovich15He}. 
Angular momentum conservation yields 
\begin{equation}\label{eq:Lfi}
 L_i= L_f \ ,
\end{equation}
where
\begin{equation}\label{eq:Lf}
  L_f=I_{s+p}\Omega_{s+p} \ .
\end{equation}
	In this case the total angular momentum of the system is the star's spin rate $\Omega_{s+p}$, where the $s+p$ denotes the final star and planet object. We solve Equation (\ref{eq:Lfi}) for $\Omega_{s+p}$. Note that the angular momentum of the pertuber should not have changed during the high eccentricity migration. Moreover, even if a scattering took place and a far away planet was lost from the system, the angular momentum associated with it is orders of magnitude smaller than the ones in Equation   (\ref{eq:Li}), and thus do not affect the above analysis.
For  $m<<M$ we find that $I_s\Omega_s$ is larger or comparable to $L_{\rm orb}$, and thus, $\Omega_{s+p,L} \sim {\rm Const.}$

As can be seen from the above equations, the key parameter in determining the final spin rate for a given star is the planet's size. This is depicted in Figure \ref{fig:timeplot} where we explore a large range of companion planets' mass from $10^{-3}$~M$_{J}$ up to $10$~M$_J$ where ~M$_J$ is the mass of Jupiter. 
We adopt the following mass radius relation for the planet  
\begin{equation}\label{eq:Rlation}
    \frac{m}{M_{\oplus}} = \left(\frac{r}{R_{\oplus}}\right)^{2.06} \ ,
\end{equation}
\citep[e.g.,][]{Lissauer+11}, where subscript $\oplus$ denotes Earth's value. {We also test a simple relation for which $m\sim 3 r$ \citep[e.g.,][]{Marcy+14}, and find consistent results.  }

In Figure \ref{fig:timeplot}, as a proof-of-concept, we initially consider three representative stellar masses, $0.8, 1$ and $1.2$~M$_\odot$ (red, blue and green lines, respectively in Figure \ref{fig:timeplot}). 
We evolve their spin-period and let each star consume a planet. We consider two merger times, one is after $100$~Myr, and the other is after $600$~Myr. In Figure \ref{fig:timeplot} we show the fractional change of the spin-period, specifically, $(P_{\Omega,i}-P_{\Omega,f})/P_{\Omega,i}$. Note that $P_{\Omega,i}>P_{\Omega,f}$ and we show the absolute magnitude in the Figure for illustrative purposes. As depicted in Figure \ref{fig:timeplot}, a more massive planet is more likely to spin up the star. 
For example, a Jupiter-mass planet can cause a spin up of about $70\%$ compared to the spin pre-merger for a $1$M$_\odot$ star. For a $1.2$~M$_\odot$, the change in spin is $20\%$ which is much less than the smaller mass star. On the other hand, an Earth-mass planet yields an insignificant change to the spin-period for any star from $0.8$M$_\odot$ to $1.2$~M$_\odot$ after $100$~Myr and $600$~Myr. Below we will examine a mass range between $0.6-1.8$~M$_\odot$.

\subsection{Energy arguments}\label{sec:energy}

{In some cases, planets that plunge into their star, may be completely consumed by their star without any heat or radiation loss for the system (unlike systems for which a dusty disk is formed e.g., \citet{Metzger+17} or increase their luminosity due to the engulfment of a planet e.g., \citet{Metzger+12};\citet{MacLeod+18}). In this case, we can assume total energy conservation, }and thus 
before the merger, it can be written as: 
\begin{equation}\label{eq:E}
    E_i \sim -\frac{GMm}{2a}- \frac{GM^2}{R} - \frac{Gm^2}{r} + \frac{1}{2}I_{s} \Omega_{s,i}^2 + \frac{1}{2}I_{p}\Omega_{p,i}^2 \ .
\end{equation}
{Note that numerical factors at the order of unity that arise from the star and planets density profiles in internal energies are neglected from the above equation for simplicity.}

Energy conservation yields that the energy after the planet has been consumed by the star can be written as: 
\begin{equation}\label{eq:Ef}
    E_i=E_f \,
\end{equation}
where $E_f$ is the energy post merger, with the subscript ``f" denoting the final (post-consumption) state. The final energy state can be written as:
\begin{equation}\label{eq:Ef}
E_f = -\frac{G(M+m)^2}{R} + \frac{1}{2}I_{s+p}\Omega_{s+p}^2 \ ,
\end{equation}
  In this case the total energy of the orbit consists of the potential energy and the star's new rotational kinetic energy, which is denoted as $\Omega_{s+p}$ to indicate that it is after the star consumed the planet. We then solve for the post merger spin-period $P_{\Omega,f} = 2 \pi / \Omega_{s+p}$.

\begin{table*}
\centering
 \begin{tabular}{l || l | l | l |l } 
 \hline
 Name & Age$^{\dagger}$  & Metallicity & Period References  & Metallicity References \\ 
 &     [Myr]  &  [Fe/H]     &      &  \\
 \hline
M35 &  100      &    -0.21            &  \citet{Meibom+09}    &  \citet{Barrado+01} \\
Pleiades & 130 &  $0.03$      &\citet{Mamajek+08}  & \citet{Soderblom+09}  \\
M34 & 240 & 0.07 & \citet{James+10}  &\citet{Schuler+03} \\
M48 & 380 & 0.08 & \citet{Barnes+15} &\citet{Netopil+16} \\
Hyades & 625  & 0.4 & \citet{Mamajek+08} &\citet{Quillen+02}\\
 \hline
 \end{tabular}
  \caption{Relevant observation parameters for the open clusters used below.${\dagger}$ Note that other age estimations exists in the literature, which we refer to below. The other age estimation do not change our results, and we discuss them in the text.
  The age estimate of M35 in the literature ranges from 70-200Myr. \citet{Reimers+88} used white dwarf cooling age and estimated a range between 70-100 Myr, later \citet{Barrado+01}  with the same technique, found an age estimated 180 Myr \citet{Kalirai+03}. Other estimations place M35 at about $150~$Myr \citep[e.g.,][]{Sarrazine+00,Hippel+02,Meibom+09,Leiner+15}. We note that while the age of M35 might be older than the one we adopt here, the scatter in the bifurcation is consistent with the scatter in the Pleiades.   
  Pleiades age varies from  $100$~Myr to $180$~Myr \citep[e.g.,][]{Mamajek+08,Herbst+01,Belikov+98}. 
  M34 age is consistent with $240$~Myr \citep[e.g.,][]{Meibom+11,James+10}. 
  M48 age varies from $380$~Myr to $450$~Myr \citep[e.g.,][]{Netopil+16,Barnes+15}. 
 Finally, we found that Hyades age estimation can be as high as $750$~Myr \citep[e.g.,][]{Mamajek+08,Brandt+15}.
 }\label{table} 
\end{table*}

As  implied from Equation (\ref{eq:E}), the planet's orbital energy is much smaller than the star's internal energy, and thus it can be neglected. We adopt a high eccentricity migration, which often results in near radial planetary orbits, and set the planet's initial semi-major axis to be $5$~au.  However, from the mentioned orbital energy arguments, the calculation below is valid for a wide range initial separations. We have also confirmed that our calculations are consistent with the results, when setting the planet to be as close as $0.02$~au. In fact, as apparent from Equations (\ref{eq:E}) and (\ref{eq:Ef}) for a given stellar mass, the main parameter that affects the final stellar rotation in this scenario is the planet's mass. 

Adopting a simple mass-radius relation as before ($m\sim r^\beta$), we find that the final spin period for very small mass planets ($m<<M$) depends on the mass of the planet i.e., \begin{equation}\label{eq:OmegaE}\Omega_{s+p,E}\sim m^{1-\frac{1}{2\beta}} \ ,\end{equation} where the subscript ``E'' stands for energy argument. {In deriving Equation (\ref{eq:OmegaE}) we assumed a high eccentricity migration, which yields that the orbital energy is much smaller than the binding energy of the planet and the star. Moreover, for typical values of star and planet spin rotation, the rotational energy is smaller than their binding energy. Finally, we focus on the dependency of final rotation rate on the planet's mass. However, we note that spin rate also depends on the star's radius and mass, since we assume that these did not change in the consumption process we drop them in the above Equation.} In the inset of Figure \ref{fig:timeplot} we show the relation between the resulted spin due to the energy or angular momentum conservation. As depicted in the Figure \ref{fig:timeplot}, ${\Omega_L}/{\Omega_E}$ is almost 1 for $M_J/1000$ ($M_\oplus$) and reaches a minimum near the mass of Jupiter. 

\section{Consumption of a planet by its star}\label{sec:con}

\begin{figure*}
 \includegraphics[width=\linewidth]{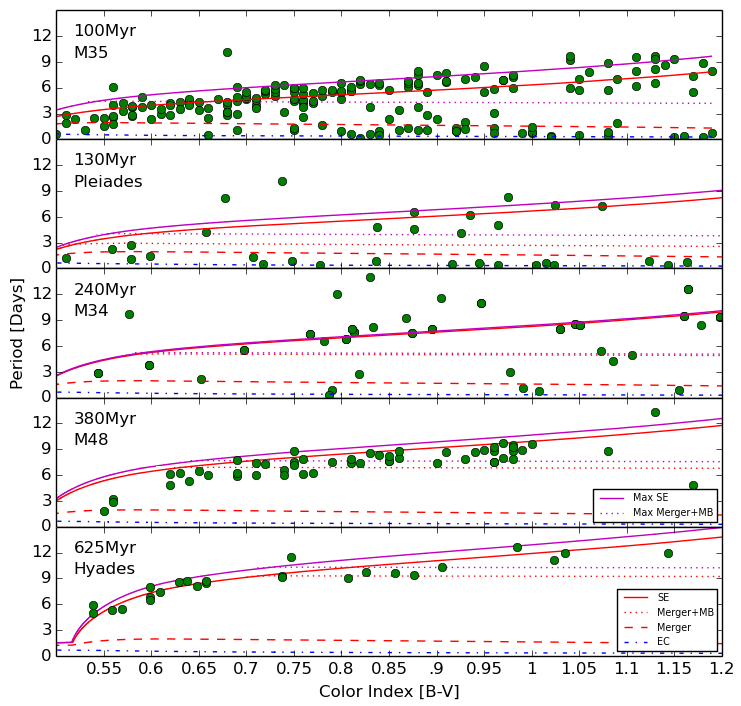}
\caption{ 
 The spin-period as a function of B-V. The solid, red, lines depict the expected period of stars for their clusters' respective ages (labeled ``SE'' following stellar evolution). The dashed lines represent the spin-period post-merger at the time of the Jupiter mass planet merger, using angular momentum arguments, labeled ``Merger'' (adopted to be consistent with the youngest cluster age of $100$~Myr).  We also consider the post-merger spin period using energy conservation arguments, labeled ``EC'', blue, short dashed-dotted line. The dotted red line depicts the expected spin-period that have undergone a merger at $100$~Myrs and subsequently went through magnetic braking process to reach their cluster present age (labeled ``Merger + MB'').  We also consider the expected spin period due to stellar evolution for the maximum published age of each cluster (see Table \ref{table}), purple, solid lines, labeled ``Max SE''. The post merger spin period (using angular momentum arguments), followed by magnetic period all the way to the maximum age estimation is shown in dotted, purple lines, labeled ``Max Merger $+$ MB''.  
The green dots are observed rotations periods of confirmed members of the above clusters adopted from \citet{Meibom+09,Mamajek+08,James+10} and \citet{Barnes+15},  see Table \ref{table}.  
\label{fig:Obs}}
\end{figure*}


We compare our model to the period distribution of stars for five open clusters that span a range of ages. Specifically, we chose (from $\approx 100$~Myr to $625$~Myr): M35, Pleiades, M34, M48, and Hyades,  see Table \ref{table} for the relevant parameters. We focus on a stellar mass range of $0.6-1.8$~M$_\odot$. As apparent from Figure \ref{fig:Obs} all the young clusters  appear to have a bifurcation in their period distribution. However, the old clusters do not  have a fast rotating population.

As illustrated in Figure \ref{fig:timeplot} the largest spin-up effect will happen if a star will merge with a massive planet. 
We adopt a Jupiter-mass planet  and explore the consequences of planet consumption on the spin-period as a function of the B-V.  For each cluster, we evolve the spin-period up to the cluster age, as explained above. As can be seen in Figure \ref{fig:Obs}, the stellar evolution model (SE) agrees with the slow rotators (long period) population\footnote{ Some of these clusters have large range of age estimations in the literature, see table \ref{table}. In Figure \ref{fig:Obs} we show that the SE  model for the maximum age estimates for these clusters. }.  We also adopt an ad-hoc consumption time consistent with the youngest cluster (M34, which is $\approx 100$~Myr). We calculate the spin-period of the star after a merger with a Jupiter sized planet has taken place, using angular momentum conservation, for all clusters (dashed lines, labeled ``Merger'' for a spin up). As shown in this Figure, the resulted post-consumption spin-period is consistent with the observed fast rotators (short spin-period) stars in the young clusters (top three panels).  Based on these five cluster examples it seems that the bifurcation  is eliminated by $\approx 300$~Myr (roughly the age of M48).

We also show that the spin period from energy conservation arguments (see Section \ref{sec:energy}), labeled ``EC'', agrees with the short period stellar population in young clusters. In fact the shortest spin periods seem to be in a better agreement with the energy conservation arguments than with the angular momentum arguments  period predictions. This suggests that near radial orbits during high eccentricity migration \citep[e.g.,][]{Naoz16} {\bf are common.} 

{It is worth noting that, for a given mass, the initial spin of star plays an insignificant role in determining post consumption spin, for both the energy and angular momentum approaches. While the initial spin of the star is an important factor for the magnetic braking and the stellar evolution process, the main contributor for the {\it post consumption} stellar spin is the size of the planet.}  Thus, taking h Persei young open cluster as a birth population, or even setting the consumption time to be $13$~Myr, does not change our results.  However, if consumption would have taken place after $13$~Myr, then our scenario predicts a clear bifurcation signature which does not exists in h Persei \citep[see, for example, figure 10 in][]{Moraux+13}. Thus, motivated with observation, and consistent with theoretical arguments \citep[e.g.,][]{Naoz+12bin,Stephan+17}, we are set the conniption time at $100$~Myr, which is roughly the age of M34, where the earliest bifurcation of rotation
periods is observed. 

	Many of the clusters are older than the consumption time, and one can expect that post-merger stars will continue spin down due to magnetic braking. This process may be different than the nominal magnetic braking as the new mass might not be evenly distributed or the metallicity and magnetic field may change. Nonetheless, for simplicity we only follow the regular magnetic braking evolution (as described above, labeled ``Merger $+$ MB''), and caution that this treatment is incomplete, as it does not include the stellar evolution complexities that SSE considers. This estimation should only be used as an order of magnitude approximation. The result is shown as a dashed-dot line in Figure \ref{fig:Obs}. And can be seen to perhaps match the scatter in those plots. 

We note that the stellar evolution is calculated in term of the effective temperature. We then convert the theoretical effective temperature to B-V colors following \citet{Fukugita+00} fit equation. This fit depends on the metallicity of the cluster. 

\begin{figure}
  \centering
    \includegraphics[width=\linewidth]{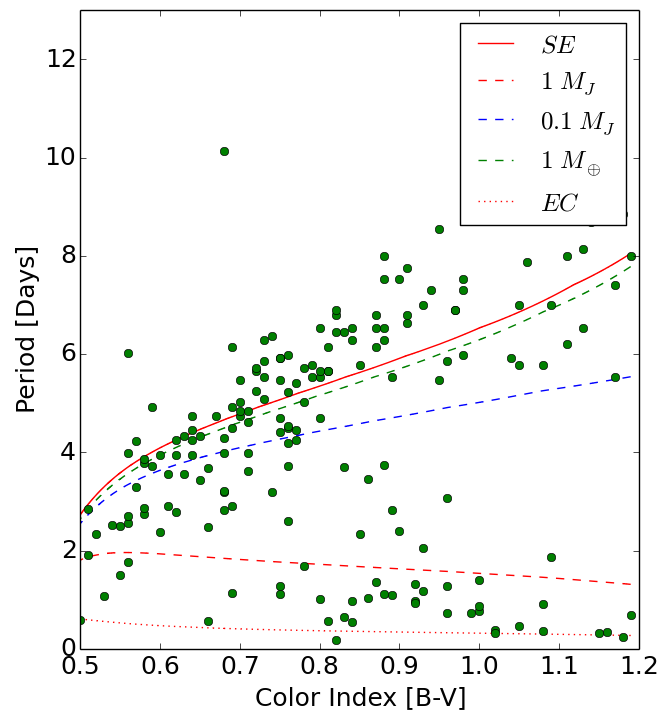}
    \caption{The spin-period as a function of the B-V values for M~35 open cluster. The solid line depicts the expected spin-period as a result of magnetic braking spin evolution after $100$~Myr, while the dashed lines represent the spin-period post-merger of a planet with mass of  $M_J$, $M_J/10$ and $M_\oplus$ (from bottom to top)f. The dotted line represents a merger of $1$~M$_J$ mass planet based on energy conservation (EC).}\label{fig:M35}
\end{figure}

As shown in Figure  \ref{fig:Obs} a Jupiter-mass planet accreting onto a star at early times ($\approx 100$~Myr) is consistent with spin up of these stars. As the cluster grows old the star spin down and approaches the nominal stellar evolution time. 

As mentioned, some of these open clusters have a range of published ages (see Table \ref{table}).  In Figure \ref{fig:Obs} we also consider the maximum age estimation, and show that our models are consistent with the observed scatter in spin period. In particular, we show the spin period as a result of the stellar evolution model for its maximum age estimate (labeled ``Max SE''), purple lines). We also calculate the spin down of a star post planet consumption, using angular momentum conservation, (purple dotted line, labeled ``Max Merger $+$ MB''), which also agrees with the observed cluster. Thus, we conclude that range of age estimation does not alter our results.

We also speculate that accretion of various planetary masses may account for the scatter in the spin-period observed for young clusters. This is illustrated in Figure \ref{fig:M35}, where we calculated the post merger  spin-period of a star after the consumption of  $M_J/10$ and an Earth mass ($M_\oplus$) planet. As depicted in Figure 3, the scatter in the young cluster is consistent with accretion of various planetary masses.

\section{Discussion}\label{sec:Diss}

We presented a proof-of-concept calculation that shows that the observed  spin-periods bifurcation in young clusters is consistent with stars that consumed a planet. We considered angular momentum conservation arguments and showed that consumption of a planet can significantly spin up the star (lowering the spin-period). Energy conservation arguments (where a planet is being plunged almost radially into the star) give consistent results. Both angular momentum and energy arguments yield a faster rotation after a consumption of a massive planet, like a Jupiter-mass one.

One of the predictions from high eccentricity migration is that planets will end up consumed by their parent star in the range of $10-100$~Myrs  \citep[e.g.,][]{Rasio+96,Sourav+08,Guillochon+11,Naoz+12bin,Li+13,Rice15,Valsecchi+14Edge,Petrovich15He,Petrovich152p,Stephan+17}. {In some cases, the perturber can yield a plunging time that can goes up to $1$~Gyr \citep[e.g.,][]{Anderson+16}}. Observationally, many giant planets seems to exist on  decaying orbits, or on orbits that are interior to their Roche-limit \citep[][Figure 1]{Jackson+17}, consistent with high eccentricity migration. Thus, planets plunging into a star may be a generic feature of planetary evolution. While the details of a planet accreting onto a star may be complicated \citep[i.e., either disrupting and forming a disk or simply colliding][]{Dosopoulou+17}, our calculations focus only on the consequences of planet consumption and hence we used energy and angular momentum conservation arguments, and we are independent of the details of the merger process.

We calculated the effects of a planetary merger on the host stellar spin, using magnetic braking, and angular and energy conservation arguments. We adopted an ad-hoc consumption time consistent with the youngest cluster ($\approx 100$~Myr) The planet's orbital angular momentum is absorbed by the star, causing the star to spin up. Similarly, energy conservation arguments yield that the planet's binding energy is being absorbed by the star, causing the star's spins up. We note that during the planet consumption process energy might not be conserved \citep[e.g.,][]{Metzger+17}, but angular momentum should be conserved\footnote{Note that the angular momentum that is carried out by the planet mass loss during the merger is significantly smaller than the orbital angular momentum. }. Note that some planets may plunge directly into their host star, consistent with minimum energy loss.
 The consumption of a more massive planet (i.e., Jupiter mass) will cause a more significant spin up (as depicted in Figure \ref{fig:timeplot})
 
We compared our calculations with the observed spin period of stars in five open clusters. 
	We find that the observed stellar rotation period bifurcation in young clusters is consistent with the spin up due to a merger of a Jupiter-mass planet  (see, Figure \ref{fig:Obs}). The agreement of our calculations with observations suggests that dynamical planetary accretion onto their host stars is a common characteristic in high eccentricity planetary-systems evolution \citep[as predicted by theoretical models, e.g.,][]{Naoz11,Guillochon+11,Naoz+12bin,Li+13,Valsecchi+14Edge,Rice15,Storch+15,Munoz+15,Petrovich15He,Petrovich152p,Stephan+17}.

	The observations presented in Figure \ref{fig:Obs} show a reduction in  obs the bifurcation strength with time. In other words, as the cluster ages, there are fewer fast rotators. 
Motivated by the observations, we adopted a merger time of $100$~Myr which is the youngest cluster considered. Interestingly, this merger time also approximately consistent with the expected merger time from high eccentricity migration simulations \citep[e.g.,][]{Stephan+17}. 

Note, that is was also suggested that disk migration will, in some cases, result in plunging Earth and super Earth type planets onto their host stars \citep[e.g.,][]{Batygin+15}\footnote{High eccentricity migration can also result in plunging Earth-size planets onto their host star \citep{Rice15}. }. The consumption of these planets yield a smaller change in the spin-period of stars (as shown in Figures \ref{fig:timeplot} and \ref{fig:M35}). On the other-hand, consumption of smaller planets, or different consumption times, may account to some of the observed scatter in the fast rotators' typical spin-period value (as depicted in Figure \ref{fig:Obs}).
{ 
We can also estimate the efficiency with which our mechanism is producing fast rotators. We note that the number below is highly uncertain and should  only be considered at the order of magnitude level. The fraction of stars that will consume a planet is defined as:
 \begin{equation}
f=f_b f_p f_{\rm merge} \ ,
\end{equation}
where $f_b$ is the fraction of stars in binary systems, $f_p$ is the fraction of stars with planets that may undergo high eccentricity migration and finally, $f_{\rm merge}$ is the fraction of planets that merged for a specific high eccentricity migration. We estimate $f_b\sim 0.5$ \cite[e.g.,][]{Raghavan+10}, though this is an underestimation since planets can also cause planets to merge with their host stars \citep[e.g.,][]{Naoz11,Petrovich152p}. The occurrences of planets is estimated roughly as $f_p\sim 0.1-0.9$ Jupiter-mass planets formed from one to few~au from their star \citep[e.g.,][]{Cumming+08,Wright+12,Bowler16} up to $f_p\sim 1$ for Neptune-mass planets (the solar system, for example, has two, Uranus and Neptune). Finally, the merger fraction is estimated as between $f_{\rm merger}\sim 0.15-0.25$ depending on the parameters of the system, and roughly independent on the mass of the planet \citep[e.g.,][]{Naoz+12bin,Petrovich15He,Anderson+16,Stephan+17}. This gives $f\sim 0.013-0.13$ The fraction of fast rotators in the cluster population is estimated from the presented data about $30\%$. This implied that about $4-42\%$ of all fast rotators are plausible to result from the process of feeding on their planets. The fraction may increase significantly if we consider the full range of planetary masses and the full high eccentricity migration scenarios (for example, allowing for a range of companions, such as planets).     
}

	If indeed the primary driver for plunging Jupiter size planets into a star is high eccentricity migration, then the calculation presented here suggests that the fast rotators are more likely to have a far away companion \citep[either a star, a planet or a brown dwarf, e.g.,][]{Rasio+96,Naoz11,Naoz+12bin,Tutukov+12,Petrovich152p,Petrovich15He,Anderson+16,Stephan+17}. This prediction may help disentangle between the scenario suggested here and magnetic origin for the fast rotators.

\acknowledgements

We thank the referee for useful and thoughtful comments. We also thank  Alexander Stephan, Jennifer van Saders, Eric Mamajek and Brad Hansen for useful discussions. SN acknowledges the partial support of the Sloan fellowship and also the partial support from the NSF through grant No.~AST-173916. ES appreciates support from the NASA Origins of Solar System grant, NNX13AH79G / 80NSSC18K0003.

 \bibliographystyle{apj}
 \bibliography{Spin}
 

\end{document}